\documentclass{acmart}
\usepackage{booktabs}
\usepackage{tabularx}
\usepackage{float}
\usepackage{multirow}  
\usepackage{algorithm}
\usepackage{array}
\usepackage{threeparttable}
\usepackage{algpseudocode}
\usepackage{xcolor}
\usepackage{listings}
\lstdefinelanguage{json}{
    basicstyle=\ttfamily\small,
    keywords={},
    keywordstyle=\color{blue},
    commentstyle=\color{gray},
    stringstyle=\color{red},
    numbers=left,
    numberstyle=\tiny\color{gray},
    stepnumber=1,
    numbersep=8pt,
    showstringspaces=false,
    breaklines=true,
    frame=single,
    backgroundcolor=\color{white},
    literate=
        *{0}{{{\color{red!20!violet}0}}}{1}
         {1}{{{\color{red!20!violet}1}}}{1}
         {2}{{{\color{red!20!violet}2}}}{1}
         {3}{{{\color{red!20!violet}3}}}{1}
         {4}{{{\color{red!20!violet}4}}}{1}
         {5}{{{\color{red!20!violet}5}}}{1}
         {6}{{{\color{red!20!violet}6}}}{1}
         {7}{{{\color{red!20!violet}7}}}{1}
         {8}{{{\color{red!20!violet}8}}}{1}
         {9}{{{\color{red!20!violet}9}}}{1}
}

\definecolor{background}{RGB}{240,240,240}
\AtBeginDocument{%
  }

\setcopyright{acmlicensed}
\copyrightyear{2018}
\acmYear{2018}
\acmDOI{XXXXXXX.XXXXXXX}
\acmConference[Conference acronym 'XX]{Make sure to enter the correct
  conference title from your rights confirmation email}{June 03--05,
  2018}{Woodstock, NY}
\acmISBN{978-1-4503-XXXX-X/2018/06}

\begin{document}

\title{Enhancing Smart Contract Vulnerability Detection in DApps Leveraging Fine-Tuned LLM}

\author{Jiuyang Bu}
\affiliation{
  \institution{School of Cyberspace Security, Hainan University}
  \city{Haikou}
  \country{China}
  \postcode{570228}
}

\author{Wenkai Li}
\affiliation{
  \institution{School of Cyberspace Security, Hainan University}
  \city{Haikou}
  \country{China}
  \postcode{570228}
}

\author{Zongwei Li}
\affiliation{
  \institution{School of Cyberspace Security, Hainan University}
  \city{Haikou}
  \country{China}
  \postcode{570228}
}

\author{Zeng Zhang}
\affiliation{
  \institution{School of Cyberspace Security, Hainan University}
  \city{Haikou}
  \country{China}
  \postcode{570228}
}

\author{Xiaoqi Li}
\affiliation{
  \institution{School of Cyberspace Security, Hainan University}
  \city{Haikou}
  \country{China}
  \postcode{570228}
}
\email{csxqli@ieee.org}
\renewcommand{\shortauthors}{Jiuyang Bu et al.}

\begin{abstract}
Decentralized applications (DApps) face significant security risks due to vulnerabilities in smart contracts \cite{DurieuxEtAl2020ICSE, slowmist}, with traditional detection methods struggling to address emerging and machine-unauditable flaws. This paper proposes a novel approach leveraging fine-tuned Large Language Models (LLMs) to enhance smart contract vulnerability detection \cite{chen2025chatgpt}. We introduce a comprehensive dataset of 215 real-world DApp projects (4,998 contracts), including hard-to-detect logical errors like token price manipulation, addressing the limitations of existing simplified benchmarks \cite{10486822}. By fine-tuning LLMs (Llama3-8B and Qwen2-7B) with Full-Parameter Fine-Tuning (FFT) and Low-Rank Adaptation (LoRA) \cite{10480574}, our method achieves superior performance, attaining an F1-score of 0.83 with FFT and data augmentation via Random Over Sampling (ROS). Comparative experiments demonstrate significant improvements over prompt-based LLMs and state-of-the-art tools. Notably, the approach excels in detecting non-machine-auditable vulnerabilities, achieving 0.97 precision and 0.68 recall for price manipulation flaws. The results underscore the effectiveness of domain-specific LLM fine-tuning and data augmentation in addressing real-world DApp security challenges, offering a robust solution for blockchain ecosystem protection.
\end{abstract}




\maketitle

\section{Introduction}
Compared to single traditional smart contracts, DApps represent a new blockchain application paradigm composed of a series of smart contracts. DApps have been widely adopted in various fields such as gaming and finance. Since smart contracts may inherently contain vulnerabilities, they inevitably introduce security risks to decentralized applications \cite{Tolmach_2021_Survey, Dwivedi_2021_Legally, Zhuang_2020_Smart}. Analyzing smart contract security is a necessary approach to enhance blockchain system safety, with vulnerability identification being the primary step in DApp security analysis \cite{Tchakounte_2022_smart, Feist_2019_Slither}. We argue that the smart contract vulnerability detection methods presented in this chapter hold significant importance for improving the security and management of decentralized applications.

LLMs have demonstrated remarkable capabilities in multiple software engineering tasks including code generation, test generation, and code summarization \cite{Ajienka_2020_empirical, Ghaleb_2023_AChecker, 10.1145/3699597}. Leveraging their versatility and innovative feature of predicting outcomes through plain-text explanations, LLMs are gradually emerging as an important research direction for security vulnerability detection \cite{AlDebeyan_2022_Improving, Wang_2022_Unified, luo2024scvhunter}. With the continuous development of blockchain technology, smart contracts are increasingly managing substantial digital assets. Attackers exploit security vulnerabilities in smart contracts to obtain substantial illicit profits \cite{Asudani_2023_Impact, Church_2017_Word2Vec, torres2019art}. According to Slowmist's statistics, blockchain security witnessed 466 hacking incidents in 2023, with vulnerabilities causing approximately \$2.49 billion in losses \cite{slowmist}. Traditional vulnerability analysis methods based on program analysis are limited by predefined vulnerability patterns, rendering them ineffective against emerging security issues \cite{Yu_2021_Knowledge, Zhou_2023_Double, brent2020ethainter}. Consequently, innovative approaches are urgently needed to combat these new blockchain security challenges.

Recent investigations reveal that while DApps still suffer from conventional contract vulnerabilities (such as reentrancy and arithmetic vulnerabilities), nearly 80\% of contract vulnerabilities are machine-unauditable and difficult to detect manually, collectively categorized as logical errors \cite{Ganin_2015_Unsupervised, song2025silence}. Emerging studies indicate that LLMs demonstrate remarkable potential in auditing smart contract vulnerabilities, particularly excelling at detecting machine-unauditable vulnerabilities \cite{mythril, Luu_2016_Making, mueller2018smashing}.

\begin{figure}[h]
  \centering
  \includegraphics[width=\linewidth]{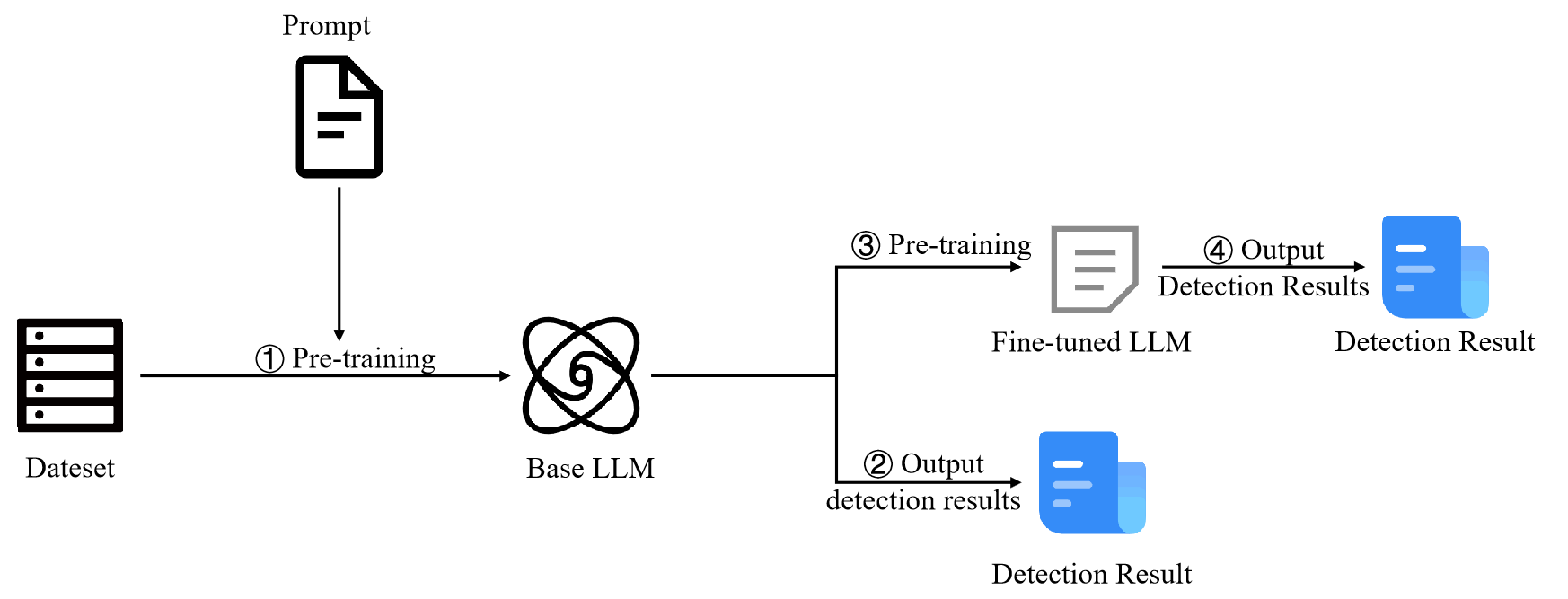}
  \caption{Basic process of LLM detection vulnerability}
  \label{fig:detect}
\end{figure}

As shown in Steps (1) and (2) of Fig. \ref{fig:detect}, current research using prompt engineering with GPT for smart contract vulnerability detection achieves only approximately 40\% accuracy \cite{tsankov2018securify, schneidewind2020ethor}. Even when employing Retrieval-Augmented Generation (RAG) to enhance LLM detection with vulnerability knowledge, the success rate remains around 60\% \cite{Torres_2021_ConFuzzius, grieco2020echidna}. This limitation exists because open-source LLMs are primarily pretrained on general code corpora without specific adaptation to Solidity, the programming language of smart contracts. As illustrated in Fig. \ref{fig:detect}, this chapter proposes Step (3) to optimize and adjust the pretrained LLM, ultimately obtaining a fine-tuned LLM specifically designed for smart contract vulnerability detection tasks \cite{Mossberg_2019_Manticore, wustholz2020harvey}.

\section{Background}
\label{sec:background}

Decentralized applications (DApps) have revolutionized blockchain ecosystems by enabling trustless transactions through smart contracts. However, the immutable nature of blockchain amplifies the consequences of vulnerabilities in these contracts \cite{brent2018vandal, Veloso__Conkas}. Traditional vulnerability detection methods, including static analysis \cite{li2024cobra} and state transition verification \cite{li2024stateguard}, have shown limitations in addressing complex vulnerabilities, particularly logical flaws in financial mechanisms like token price manipulation \cite{torres2018osiris, pakala}. For instance, while tools such as COBRA \cite{li2024cobra} employ interaction-aware bytecode analysis to detect reentrancy and integer overflow, they struggle with multi-contract dependencies and semantic-level vulnerabilities prevalent in real-world DApps \cite{li2024defitail, ma2025uncovering}.

Recent studies highlight the growing sophistication of attacks targeting decentralized ecosystems, including NFT wash trading \cite{niu2024unveiling} and cross-contract exploits \cite{wang2023smart, 10700860}. The Solana NFT ecosystem analysis \cite{kong2024characterizing} further reveals that 23\% of vulnerabilities stem from non-machine-auditable flaws, underscoring the need for advanced detection paradigms \cite{zhen2024gnn, shang2025cegt}. Existing benchmarks often oversimplify contract structures, failing to capture the complexity of production-level projects \cite{li2023overview, 10.1145/3543507.3583367}. This data gap hinders the training of AI-driven detectors, despite promising results from LLM-based approaches in related tasks like contract summarization \cite{mao2024scla}.

The integration of AI and blockchain has emerged as a promising direction \cite{li2023overview, LI2026122645}, with recent works like SCALM \cite{li2025scalm} demonstrating LLMs' potential in identifying code smells. However, prompt-based LLM methods exhibit inconsistent performance for vulnerability detection due to domain-specific semantic gaps \cite{li2024detecting, liu2025detecting}. Fine-tuning strategies, particularly those addressing data imbalance through techniques like Random Over Sampling (ROS), have shown efficacy in analogous domains such as sandwich attack detection \cite{10707457, ZHU2021107920}. Meanwhile, cross-ecosystem studies \cite{li2020characterizing,li2017discovering} emphasize the critical role of transaction graph analysis in understanding adversarial patterns, suggesting untapped potential for hybrid approaches combining LLMs and behavioral analysis \cite{10646885}.

These challenges motivate our work to bridge three critical gaps: (1) the lack of comprehensive datasets reflecting real-world DApp complexity, (2) the limited adaptability of general-purpose LLMs to blockchain security contexts, and (3) the absence of robust methods for detecting logical vulnerabilities that evade conventional program analysis \cite{zhang2025bounded, li2024detecting, chen2024improving}. Our approach builds upon foundational insights from state derailment detection \cite{li2024guardians} and hybrid analysis frameworks \cite{li2021hybrid}, extending them through domain-specific LLM fine-tuning and semantic-aware data augmentation \cite{runtimeverification}.

\begin{figure}[t]
  \centering
  \includegraphics[width=0.7\linewidth]{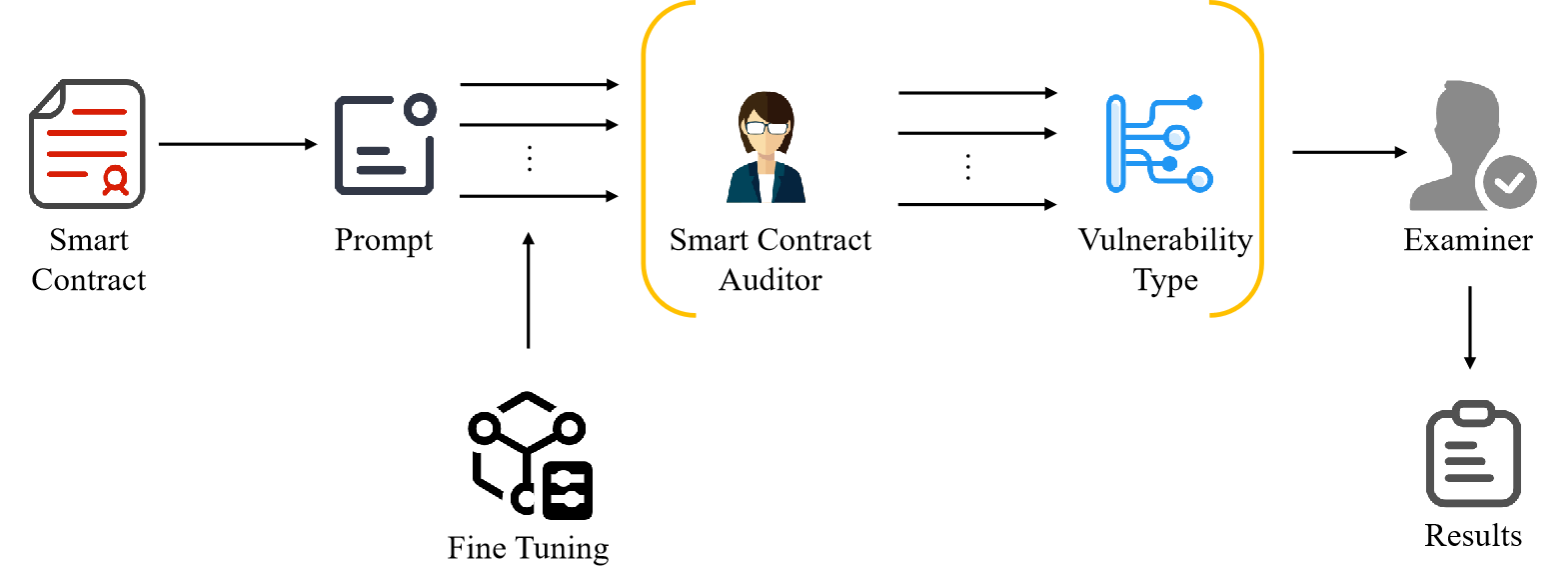}
  \caption{LLM Dual Audit-Verification for Smart Contract Vulnerabilities}
  \label{fig:audit}
\end{figure}

\section{Method}
This chapter employs a fine-tuning approach to adapt LLMs for Solidity code vulnerability detection, thereby enabling intuitive smart contract auditing and verification \cite{chen2025chatgpt, 10.1145/3699597}. As illustrated in Fig. \ref{fig:audit}, the LLM assumes two roles—Smart Contract Auditor and Verifier—activated through prompt instructions.

\textbf{Smart Contract Auditor:} Responsible for auditing smart contract code. By training the LLM with code containing known vulnerabilities, the auditor can identify potential security issues and generate identified vulnerabilities along with relevant reasoning.

\textbf{Verifier:} Responsible for further verifying the types of identified vulnerabilities to ensure the accuracy and rigor of audit results.

\subsection{Decentralized Application Contract Data Collection}
Although many research works have published datasets for smart contract vulnerability detection, most of these datasets contain contract codes with only 40 lines or fewer, far below the scale of real-world DApp projects \cite{10486822, DurieuxEtAl2020ICSE}. Vulnerability detection tools that perform well on such simplified contracts cannot guarantee consistent performance in real-world complex DApp scenarios. To address the lack of representativeness in existing datasets, this chapter collects audited contract projects from Code4rena and Slowmist platforms and implements a Python tool called \texttt{SmartCollect} to recover a complete DApp project's dependency contracts, including public libraries and external contracts.

To better obtain complete DApp projects, the tool scans all \texttt{.sol} files in the project and excludes irrelevant configuration files such as \texttt{.js} files. For each \texttt{.sol} file, the tool retrieves dependency contracts based on import statements (e.g., \texttt{import "./aave/ILendingPool.sol";}) and returns a list of imported file paths in Solidity files. It scans the input directory for \texttt{.sol} files referenced in imports and identifies missing \texttt{.sol} files. After identifying dependency contracts for each DApp, missing library contracts (e.g., ERC20, often from OpenZeppelin) are manually collected from GitHub and automatically integrated into the DApp project to ensure completeness.

Using \texttt{SmartCollect}, 215 DApp projects comprising 4,998 smart contracts were collected. These contracts contain vulnerabilities summarized in the SWC Registry. While vulnerabilities like reentrancy and arithmetic issues persist across Solidity versions, others such as unprotected self-destruct and default function visibility have been mitigated in newer versions. Thus, this chapter focuses on reentrancy, arithmetic vulnerabilities, and timestamp dependence. Additionally, to evaluate the LLM's performance on logical errors, 23 contracts with token price manipulation vulnerabilities were collected. The dataset was organized as shown in Fig.\ref{fig:data} and saved in \texttt{.json} format.


\begin{figure}[ht]
\begin{lstlisting}[language=json]
{
    "messages": [
        {
            "role": "system",
            "content": "You are a smart contract auditor.Review the following smart contract code in detail and identify vulnerabilities type within it."
        },
        {
            "role": "user",
            "content": "Source code&Vulnerability Related Description"
        },
        {
            "role": "assistant",
            "content": "Vulnerability Type"
        }
    ]
}
\end{lstlisting}
\caption{Dialogue Format for LLM-Based Smart Contract Audits} 
\label{fig:data}      
\end{figure}

\subsection{Data Augmentation}
A prominent and practical challenge in vulnerability detection is the issue of data imbalance \cite{ZHU2021107920}. In real-world projects, the ratio of vulnerable to non-vulnerable cases is highly skewed. According to existing research, the performance of state-of-the-art deep learning-based code vulnerability detection methods deteriorates significantly when applied to imbalanced real-world data. To address the data imbalance problem in our dataset, this chapter proposes Random Over Sampling (ROS) for dataset optimization.

Over-sampling balances the dataset by increasing the number of samples in the minority class. For instance, the ROS method randomly selects samples from the minority class and duplicates them to augment the training data. This process enhances the representation of minority classes, thereby mitigating the impact of data imbalance on model performance. ROS has been demonstrated to be robust.

\subsection{Prompt Engineering Design}
\begin{figure}[h]
  \centering
  \includegraphics[width=0.5\linewidth]{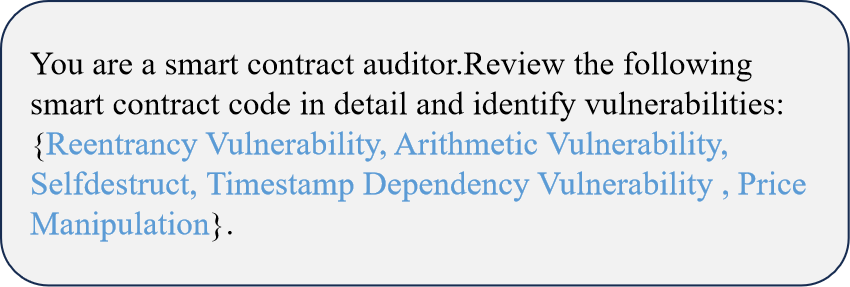}
  \caption{Basic Prompts}
  \label{fig:prompt1}
  \vspace{-1em}
\end{figure}

Current research indicates that prompt engineering significantly influences the effectiveness of smart contract vulnerability detection. In this chapter, we define prompts as scenarios tailored to specific roles and requirements. As shown in Fig. \ref{fig:prompt1}, our prompts include clear role definitions, prior knowledge, and response format specifications. We instruct the LLM to act as a smart contract auditor, identifying four types of vulnerabilities in the provided contracts.

\begin{figure}[h]
  \centering
  \includegraphics[width=0.5\linewidth]{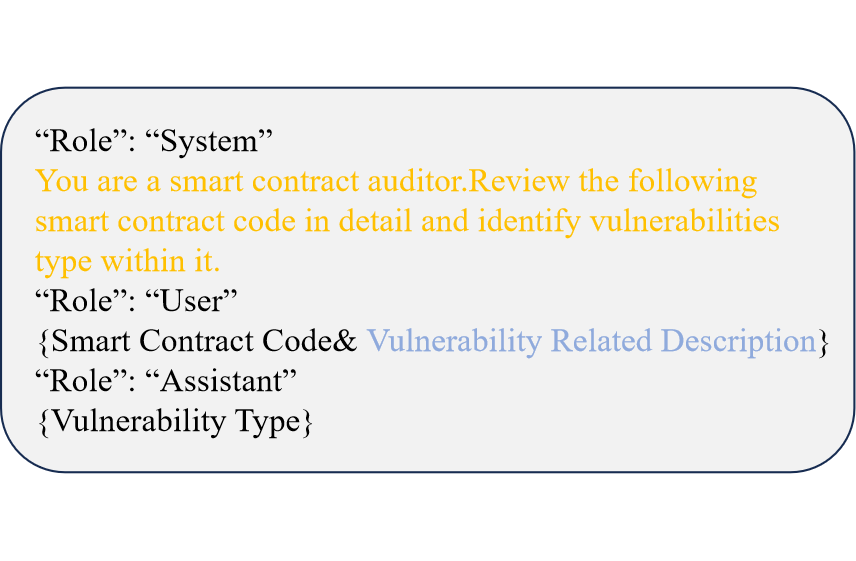}
  \caption{Code Detection Prompts}
  \label{fig:prompt2}
   \vspace{-1em}
\end{figure}

To better evaluate the LLM's capability in detecting smart contract vulnerabilities and analyzing code, we compare its detection results with audit reports as benchmarks. Following the Chain-of-Thought (CoT) approach, we designed a prompt (Fig. \ref{fig:prompt2}) that requires the LLM to emulate an auditor's workflow. The auditing process involves thoroughly understanding the smart contract's context and objectives, followed by comprehensive analysis of the logic underlying potential vulnerabilities. The prompt design incorporates role definition, CoT reasoning, requirement specifications, and response format. Notably, the ``Vulnerability Related Description'' serves as the intermediate reasoning step.

\subsection{Fine-Tuning Process}

\subsubsection{Full-Parameter Fine-Tuning}
The classical fine-tuning method primarily refers to Full-parameter Fine-Tuning (FFT), which is considered more powerful than LoRA fine-tuning in terms of effectiveness. It involves updating all parameters of the pre-trained model to achieve optimal performance on new tasks. Before performing full-parameter fine-tuning, the model typically undergoes a pre-training phase. After obtaining the base pre-trained model, we further train it using a specific dataset containing smart contract vulnerabilities to adapt it to the vulnerability detection task.

The core principle of full-parameter fine-tuning lies in retraining all model parameters for the smart contract vulnerability detection task based on task-specific data to achieve optimal performance. During FFT, all weights $W$ of the pre-trained model are treated as learnable parameters and updated via gradient descent.

Specifically, given a task dataset $\mathcal{D} = \{(x_i,y_i)\}$, full-parameter fine-tuning aims to minimize the task-specific loss function $\mathcal{L}$, where $x_i$ is the input and $y_i$ is the output label. Through backpropagation, we compute the gradient of the loss function with respect to the model parameters $W$, as shown in Equation (\ref{eq:5-1}):

\begin{equation}
W \leftarrow W - \beta\nabla_W\mathcal{L}(W;\mathcal{D})
\label{eq:5-1}
\end{equation}

where $\beta$ is the learning rate, and $\nabla_W\mathcal{L}(W;\mathcal{D})$ represents the gradient of the loss function with respect to parameters $W$. During FFT, model parameters undergo significant updates to adapt to the specific task requirements.

\subsubsection{LoRA Fine-Tuning}
LoRA (Low-Rank Adaptation of Large Language Models) is a lightweight fine-tuning method designed to enhance the adaptability and efficiency of LLMs \cite{10480574}, particularly for task-specific customization under limited resources and data conditions. The core idea of LoRA is to efficiently adjust the weights of pre-trained models by introducing low-rank matrices, avoiding large-scale retraining of the entire model and significantly reducing training time for smart contract vulnerability detection models.

\begin{figure}[h]
  \centering
  \includegraphics[width=0.7\linewidth]{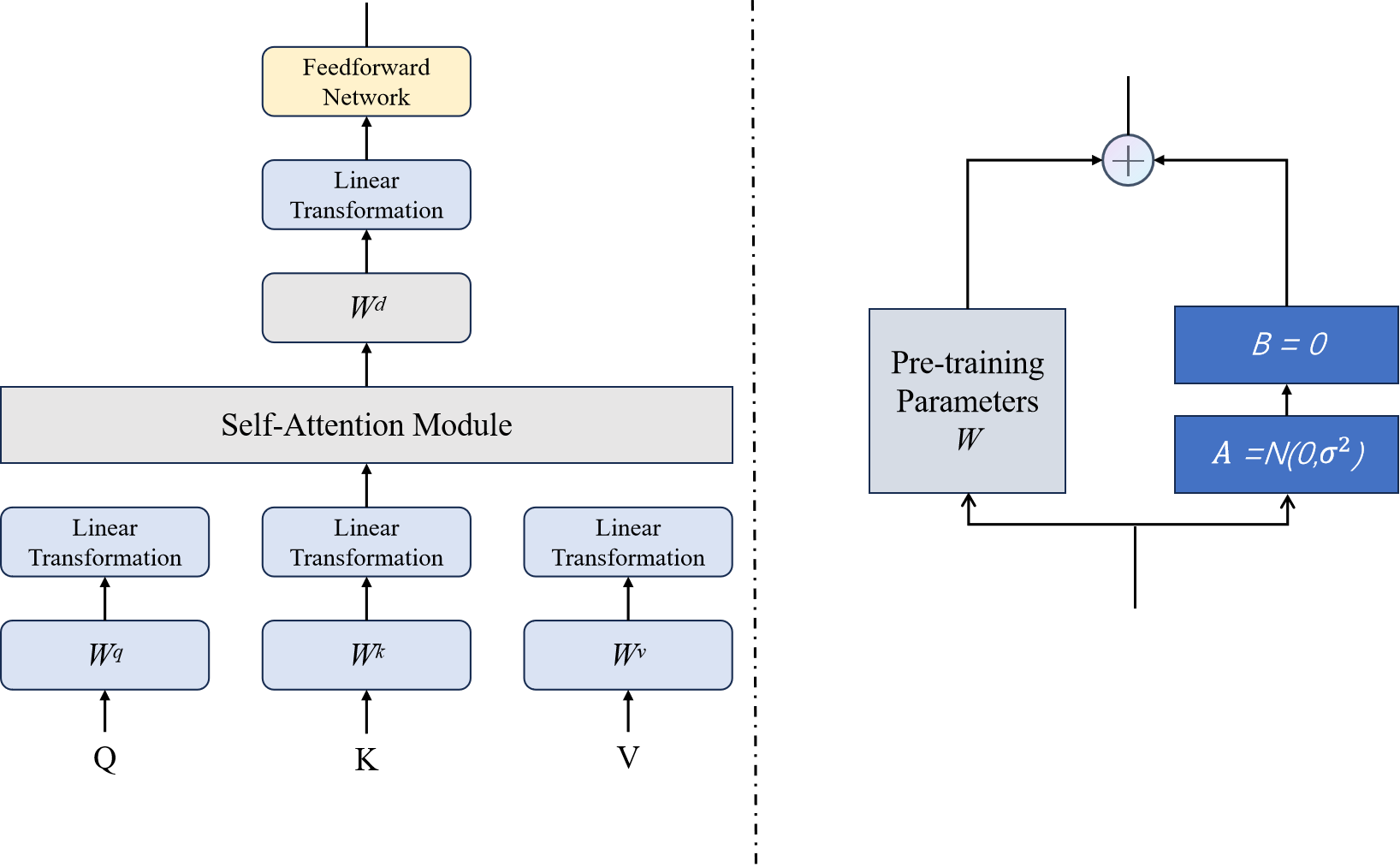}
  \caption{LoRA: Parameter-Efficient Attention Layer Adaptation}
  \label{fig:LORA}
   \vspace{-1em}
\end{figure}

As shown in Fig. \ref{fig:LORA}: During LoRA fine-tuning, matrix $A$ is initialized with random Gaussian values while $B$ is zero-initialized, resulting in $\Delta W$ being initialized to 0. Only $W_q$, $W_k$, and $W_v$ weights are adjusted.

In deep neural networks, particularly in autoregressive large language models, weight matrices typically have high dimensionality. For a given fully-connected layer, its weight matrix $W \in \mathbb{R}^{d \times k}$, where $d$ is the input dimension and $k$ is the output dimension. The key idea of LoRA is to approximate weight updates using low-rank matrix decomposition, representing the updated weight matrix as the product of two low-rank matrices, as shown in Equation (\ref{eq:5-2}):

\begin{equation}
\Delta W = AB^T
\label{eq:5-2}
\end{equation}

where $A \in \mathbb{R}^{d \times r}$ and $B \in \mathbb{R}^{k \times r}$ represent low-rank matrices, with $r \ll \min(d, k)$ being the rank size. Through this decomposition, the model only needs to learn and store these two low-rank matrices ($A$ and $B$) instead of updating the entire weight matrix $W$. Thus, LoRA achieves model fine-tuning while dramatically reducing parameters.

During model fine-tuning, the weight matrix $W$ is updated through low-rank approximation, as shown in Equation (\ref{eq:5-3}):

\begin{equation}
W_{new} = W_0 + \Delta W = W_0 + AB^T
\label{eq:5-3}
\end{equation}

where $W_0$ is the original weight matrix from the pre-trained language model, and $\Delta W = AB^T$ represents the low-rank update learned by LoRA. Only $A$ and $B$ are updated as trainable parameters, while $W_0$ remains frozen and does not receive gradient updates. This strategy effectively reduces the complexity of fine-tuning while avoiding the massive computational overhead required for full-parameter fine-tuning.

\section{Experimental Results and Analysis}
\subsection{Performance Evaluation}
In this section, we evaluate the performance of Fine-Tuning LLM and address the following research questions:

\begin{itemize}
\item RQ1: How does Fine-Tuning LLM perform in smart contract vulnerability detection tasks? We focus on metrics including accuracy, precision, recall, and F1-Score.
\item RQ2: Can the Fine-Tuning LLM-based approach outperform prompt-based methods in smart contract vulnerability detection?
\item RQ3: How effective is the proposed method in detecting non-machine-auditable vulnerabilities (e.g., price manipulation)?
\item RQ4: What is the effect of data augmentation on Fine-Tuning LLM performance?
\end{itemize}

\subsection{RQ1 - Performance Evaluation}
Using the dataset collected, we trained mainstream open-source LLMs (Llama3-8B and Qwen2-7B  with both FFT and LoRA fine-tuning techniques. The results are shown in Table \ref{tab:5-1}.

\begin{table}[htbp]
\centering
\caption{Performance comparison of Fine-Tuning LLM using FFT and LoRA}
\label{tab:5-1}
\begin{tabular}{lccc}
\hline
\textbf{Method} & \textbf{PRE} & \textbf{REC} & \textbf{F1} \\ \hline
Llama3-8B-FFT & 0.78 & 0.70 & 0.73 \\ 
Qwen2-7B-FFT & 0.78 & 0.64 & 0.70 \\
Qwen2-7B-LoRA & 0.74 & 0.57 & 0.64 \\ \hline
\end{tabular}
\end{table}

Without any preprocessing of smart contract source code, the fine-tuned LLMs achieved promising performance in vulnerability detection, with maximum precision of 0.78, recall of 0.70, and F1-score of 0.73 \cite{luo2024scvhunter, zhen2024gnn}. Table \ref{tab:5-1} shows that FFT significantly outperforms LoRA in both precision and recall, as LoRA's parameter freezing limits the model's learning capacity for complex vulnerability detection tasks.

\subsection{RQ2 - Impact of Fine-Tuning}
Table \ref{tab:5-2} compares the performance of Qwen2 models with and without fine-tuning.

\begin{table}[htbp]
\centering
\caption{Performance of Qwen2 without fine-tuning}
\label{tab:5-2}
\begin{tabular}{cccc}
\hline
\textbf{Method} & \textbf{PRE} & \textbf{REC} & \textbf{F1} \\
\hline
Qwen2-7B-Pormpt & 0.14 & 0.37 & 0.20 \\ \hline
\end{tabular}
\end{table}

The results demonstrate that LLMs without fine-tuning perform poorly (F1=0.20) in smart contract vulnerability detection, highlighting the importance of domain-specific fine-tuning for complex tasks \cite{chen2024improving, shang2025cegt}.

We further compared our best FFT models with state-of-the-art methods (GPTLENS and GPTSCAN) using precision metrics (Table \ref{tab:5-3}).

\begin{table}[htbp]
\centering
\caption{Comparison with state-of-the-art methods}
\label{tab:5-3}
\begin{tabular}{cc}
\hline
\textbf{Method} & \textbf{PRE} \\ \hline
GPTLENS & 0.64 \\
GPTSCAN & 0.57 \\
Llama3-8B-FFT & 0.78 \\
Qwen2-7B-FFT & 0.78 \\ \hline
\end{tabular}
\end{table}

Our FFT-LLMs (PRE=0.78) significantly outperform GPTLENS (0.64) and GPTSCAN (0.57) \cite{10646885, 10700860}.

\subsection{RQ3 - Price Manipulation Vulnerability Detection}
Table \ref{tab:5-4} shows the performance across different vulnerability types.

\begin{table}[htbp]
\centering
\begin{threeparttable}
\caption{Model Comparison on Vulnerability Detection}
\label{tab:5-4}
\begin{tabular}{@{} l l *{4}{c} @{}}
\toprule
\multirow{2}{*}{\textbf{Method}} & \multirow{2}{*}{\textbf{Metric}} & \multicolumn{4}{c}{\textbf{Evaluation Metrics}} \\
\cmidrule(l){3-6}
 & & RV & AV & TDV & PMV \\
\midrule
\multirow{2}{*}{Llama3-8B-FFT} 
 & PRE & 0.82 & 0.63 & 0.47 & 0.97 \\
 & REC & 0.63 & 0.70 & 0.98 & 0.68 \\ \hline
\addlinespace[0.2em]
\multirow{2}{*}{Qwen2-7B-FFT}
 & PRE & 0.57 & 1.00 & 0.44 & 1.00 \\
 & REC & 0.83 & 0.19 & 0.93 & 0.63 \\ \hline
\addlinespace[0.2em]
\multirow{2}{*}{Qwen2-7B-LoRA}
 & PRE & 0.66 & 0.82 & 0.22 & 0.97 \\
 & REC & 0.68 & 0.21 & 0.74 & 0.63 \\
\bottomrule
\end{tabular}
\end{threeparttable}
\end{table}

All models achieved high precision ($\geq 0.97$) and recall ($\geq 0.63$) for PMV detection, demonstrating LLMs' effectiveness against price manipulation vulnerabilities \cite{liu2025detecting, ma2025uncovering}.
The presence of multiple vulnerabilities in contracts may increase false positives, slightly reducing precision.

\subsection{RQ4 - Data Augmentation Effectiveness}
Table \ref{tab:5-5} presents the impact of ROS data augmentation.
\begin{table}[htbp]
\centering
\caption{Performance with ROS augmentation}
\label{tab:5-5}
\begin{tabular}{cccc}
\hline
\textbf{Method} & \textbf{PRE} & \textbf{REC} & \textbf{F1} \\ \hline
Llama3-8B-FFT-ROS & 0.84 & 0.79 & 0.83 \\
Qwen2-7B-FFT-ROS & 0.85 & 0.80 & 0.82 \\ \hline
\end{tabular}
\end{table}

Compared with Table \ref{tab:5-1}, ROS augmentation improved precision by 6-7\% and recall by 9-16\%, confirming its effectiveness for imbalanced data in LLM-based vulnerability detection \cite{LI2026122645}.

\section{Conclusion}
This paper proposes a smart contract vulnerability detection method based on fine-tuned LLMs \cite{chen2025chatgpt, luo2024scvhunter}, adapt LLMs through fine-tuning for Solidity code vulnerability detection. To analyze the security of smart contracts on the chain side of decentralized applications, this study collected relevant smart contracts in the context of decentralized applications. In addition to conventional contract vulnerabilities, the dataset specifically includes 23 contracts involving price manipulation vulnerabilities, which are prevalent in decentralized finance \cite{slowmist, torres2018osiris}. The effectiveness of fine-tuned LLMs in smart contract vulnerability detection is demonstrated across three metrics: precision, recall, and F1-score. The fine-tuned LLM exhibits outstanding performance in detecting price manipulation vulnerabilities, indicating that LLMs can be effectively applied to scenarios involving machine-unauditable contract vulnerabilities. This suggests that fine-tuned LLMs possess strong adaptability and generalization capabilities, enabling them to address complex and challenging-to-audit smart contract security issues \cite{chen2024improving, song2025silence}. Furthermore, this study enhances the performance of the fine-tuned LLM through data augmentation using the ROS technique.

\bibliographystyle{ACM-Reference-Format}
\bibliography{paper1}

\end{document}